\documentclass[eqsecnum,superscriptaddress,showpacs]{revtex4}
\usepackage{bm}	
\usepackage[dvips]{graphicx}
\usepackage{latexsym}
\begin{document}
\title{Path integral and Sommerfeld quantization}

\author{Mikoto Matsuda}\email{mmatsuda@niken.jp}
\affiliation{Japan Health and Medical technological college, Tokyo, Japan}

\author{Takehisa Fujita}\email{fffujita@phys.cst.nihon-u.ac.jp}
\affiliation{College of Science and Technology, Nihon University, Tokyo, Japan}

\date{\today}%

\begin{abstract}

The path integral formulation can reproduce the right energy spectrum 
of the harmonic oscillator potential, but it cannot resolve the Coulomb potential 
problem. This is because the path integral cannot properly take into account 
the boundary condition, which is due to the presence of the scattering states 
in the Coulomb potential system. On the other hand, the Sommerfeld 
quantization can reproduce the right energy spectrum of both harmonic oscillator 
and Coulomb potential cases since the boundary condition is effectively 
taken into account in this semiclassical treatment.  The basic difference between 
the two schemes should be that no constraint is imposed on the wave function 
in the path integral while the Sommerfeld quantization rule is derived 
by requiring that the state vector should be a single-valued function. 
The limitation of the semiclassical method is also clarified in terms of 
the square well and $\delta(x)$ function potential models. 

\end{abstract}


\maketitle

\section{Introduction}

Quantum field theory is the basis of modern theoretical physics and it is 
well established by now \cite{bd,sakurai,nishijima,fk}. 
If the kinematics is non-relativistic, then one obtains the equation of 
quantum mechanics  which is the Schr\"odinger equation. 
In this respect, if one solves the Schr\"odinger equation, then one can properly 
obtain the energy eigenvalue of the corresponding potential model . 

Historically, however, the energy eigenvalue is obtained without solving 
the Schr\"odinger equation, and the most interesting method is known as 
the Sommerfeld quantization rule which is the semiclassical method 
\cite{sommer1,sommer2,sommer3,sommer4}. 
In this case, the Sommerfeld quantization rule is to first assume
\begin{eqnarray}
\oint pdr =nh \label{RS1}
\end{eqnarray}
where $n$ denotes an integer and $h$ is the Planck constant. 
This method can be applied to the Coulomb potential problem of $V(r)=-{Ze^2\over r}$. 
In fact, it can be solved exactly by the Sommerfeld quantization. 
Here, the momentum $p$ is related to the energy $E$ as
\begin{eqnarray}
E={p^2\over 2m}-{Ze^2\over r} 
\end{eqnarray}
where $m$  denotes the mass of electron which is bound 
in the Coulomb potential. In this case, eq. (\ref{RS1}) can be written as
\begin{eqnarray}
\oint pdr = 2\sqrt{2m|E|} \int_0^{r_0} \sqrt{ {r_0-r\over r} } dr  =nh \label{RS2}
\end{eqnarray}
where  $r_0={Ze^2\over |E|}$ is introduced.  After the integration of 
eq.(\ref{RS2}), we obtain the energy eigenvalue $E$ as
\begin{eqnarray}
E=-{m (Ze^2)^2\over 2 \hbar^2 n^2}
\end{eqnarray}
which is a correct expression. The advantage of this method is clear in that it does 
not require the solution of the differential equation. Instead, one should make an 
integration of the space coordinates over the limited range of space. 

The similar type of advantage is also known in the path integral formulation, 
and indeed the path integral method is formulated so that the spectrum can be 
obtained by the infinitely many dimensional integration over the discretized 
space coordinates. However, the path integral method can be successfully applied 
only to the harmonic oscillator potential case. The reason of the successful 
description of the harmonic oscillator potential is clear because there is 
no scattering state in the harmonic oscillator potential system. 
On the other hand, the path integral cannot reproduce the energy spectrum 
of the Coulomb potential case. This is related to the fact that the Coulomb case 
contains  scattering states, and thus the value of $\int_{-\infty}^\infty |
\psi(x)|^2 dx$ is not necessarily unity. Therefore, the boundary condition cannot be 
properly taken into account in the path integral method, which should be an intrinsic 
defect arising from the formulation itself. 
 
In addition, we discuss the Sommerfeld quantization rule 
\cite{sommer1,sommer2,sommer3} and clarify  why it is successful for obtaining 
the energy spectrum of both the Coulomb and the harmonic oscillator potential systems. 
As we see below, the integration of the Sommerfeld quantization rule 
over coordinate space can be limited to the finite range of space. Therefore, 
the boundary condition is effectively taken into account in this treatment. 
However, the scattering problem cannot be treated in this semiclassical method, 
and the scattering states can be obtained by the WKB method which is the semiclassical 
approximation for the Schr\"odinger wave function. This WKB method is, however, not 
exact and only valid for the limited range of applications. The exact calculation of 
the scattering problem should be possible only when one solves 
the Schr\"odinger equation.

\section{Boundary condition in path integral formulation}

Here, we discuss the problem of the boundary condition how we can 
consider the physics of the boundary condition into the path integral expression.  
In fact, there is no way to include the boundary condition in the path 
integral formulation since the path integral does not make use of any information 
on the wave function. This is the basic reason why one cannot solve 
the Coulomb potential problem in the path integral calculation since 
the Coulomb case contains the scattering states in which the boundary 
condition in the scattering state is completely different from the bound 
state problem. In this section, we take the representation of $\hbar=1$.

\subsection{Path integral in quantum mechanics}

Here, we briefly describe the path integral formulation in quantum mechanics 
\cite{feyn1,feyn2,feyn3}. 
First, we define the transition amplitude $K(x',x:t)$ as
\begin{eqnarray}
K(x',x:t) = \langle x'|e^{-iHt} |  x \rangle = \sum_n \psi_n^* \, (x') \, 
 \psi_n(x) \, e^{-iE_nt}  \label{PI1}
\end{eqnarray}
where Hamiltonian $H$ is defined as
\begin{eqnarray}
H={ \hat{p}^2\over 2m} +V(x) = -{ \hbar^2\over 2m}{\partial^2\over \partial x^2} 
+V(x) 
\end{eqnarray}
and $\psi_n(x)$ is an eigenstate of the Hamiltonian $H$ 
\begin{eqnarray}
H \psi_n(x)=E_n\psi_n(x) . 
\end{eqnarray}
In this case, the path integral formulation is written as
\begin{eqnarray}
 K(x',x:t) =  \int {\cal D} x \ 
\exp \left\{ i \Delta t \sum_{i=1}^n \left({m(x_i-x_{i-1})^2\over{ 2(\Delta t)^2} }
 -V(x_i)\right)  \right\}  
\end{eqnarray}
where many dimensional integration $\int {\cal D} x$ is introduced as
\begin{eqnarray}
 \int {\cal D} x \equiv \lim_{n\rightarrow \infty} \left( {m\over 2i\pi\Delta t }
\right)^{n\over 2}  \int\limits_{-\infty}^\infty dx_1 \cdots \int 
\limits_{-\infty}^\infty dx_{n-1}  
\end{eqnarray}
which is a symbolic notation in the path integral formulation. 
Therefore, the transition amplitude can be written in terms of the path integral 
formulation as
\begin{eqnarray}
 K(x',x:t)  = \langle x'|e^{-iHt} |  x \rangle = \int {\cal D} x 
\ \exp \left( i \int_0^t L(x,\dot{x}) dt \right)  \label{PI2}
\end{eqnarray}
where $L(x,\dot{x}) $ is defined as
\begin{eqnarray}
L(x,\dot{x}) =  {1\over 2}\,m {\dot x}^2 -V(x) 
\end{eqnarray}
which is the Lagrangian of the corresponding classical system. 

\subsection{Boundary condition in path integral formulation}

Now a question may arise as to whether we may take into account the boundary 
condition for the wave function $\psi(x)$ in order to solve a bound state 
problem. That is, we must take into account the condition  
\begin{eqnarray}
\psi(\pm \infty) =0  \label{BC1}
\end{eqnarray}
in the path integral expression of eq.(\ref{PI1}) and eq.(\ref{PI2}) in some way or 
the other. However, one sees that the boundary condition cannot be taken into account 
in the path integral formulation in which any condition on the wave function 
never appears in this formulation. This will be shown below more in detail. 

\subsection{Boundary condition in the harmonic oscillator potential system}

There is one example which is solved exactly in terms of the path integral 
formulation, that is, the harmonic oscillator potential system. 
Why is it that the harmonic oscillator potential system can be 
successfully solved ? 
The reason is, of course, simple in that the harmonic oscillator potential 
system does not contain any scattering states, and in fact, all the states 
are bound. Therefore, it is clear that 
\begin{eqnarray}
\int_{-\infty}^\infty |\psi_n(x)|^2 dx =1
\end{eqnarray}
is always guaranteed. Thus, we obtain from eq.(\ref{PI1})
\begin{eqnarray}
 \int K(x,x:t) dx = \int \langle x|e^{-iHt} |  x \rangle dx = 
\int dx \sum_n |\psi_n(x)|^2 e^{-iE_nt} =\sum_n  e^{-iE_nt}. \label{PI3}
\end{eqnarray}
In fact, we obtain the $K(x',x:t)$ for the harmonic oscillator potential system as
\begin{eqnarray}
K(x',x:t) = \sqrt{ m\omega\over 2i\pi \sin \omega t} 
\exp \left\{i{m\omega\over{ 2} }\left[ ({x'}^2+x^2 )\cot \omega t -
{2x'x\over \sin \omega t} \right] \right\}
\end{eqnarray}
and therefore
\begin{eqnarray}
 \int K(x,x:t) dx = 
\int_{-\infty}^\infty dx \sqrt{ m\omega\over 2i\pi \sin \omega t} 
e^{-im\omega x^2 \tan {\omega t \over 2}}={1\over 2i\sin  {\omega t \over 2} } 
= \sum_{n=0}^\infty  e^{-iE_nt} .
\label{PI4}
\end{eqnarray}
Here, we make use of the following equation
\begin{eqnarray}
{1\over 2i\sin  {\omega t \over 2} }= {e^{-{i\over 2} \omega t } 
\over {1-e^{-i\omega t} } } =\sum_{n=0}^\infty e^{-i\omega t (n+{1\over 2})}
\end{eqnarray}
and thus find
\begin{eqnarray}
 E_n=\omega \left(n+{1\over 2} \right) 
\end{eqnarray}
which is a correct energy for the harmonic oscillator potential system.

\subsection{Boundary condition with scattering states}

The harmonic oscillator potential is a very special and exceptional case in quantum 
mechanics, and it is, of course, unrealistic and cannot be applied 
to the description of nature. The reason why we often find the 
harmonic oscillator potential is simple. If we treat many body systems and 
make some approximations to obtain an effective one body potential, then 
we always obtain the harmonic oscillator potential near the minimum point. 
Namely, if we treat it in terms of the small vibration around the minimum 
point $x=x_0$ of the complicated potential, then we find
\begin{eqnarray}
V(x) = V(x_0) + {1\over 2} V''(x_0) (x-x_0)^2 +\cdots .
\end{eqnarray}
and here the first term is a constant and the second term is just the harmonic 
oscillator potential. However, it is clear that this is only valid for the small 
vibrations. 

The complicated realistic potential should contain the scattering 
states since the potential $V(x)$ should vanish at the large value of $x$. 
For any systems with scattering states $\psi_{s}(x)$, the condition 
\begin{eqnarray}
\int |\psi_{s}(x)|^2 dx < \infty  
\end{eqnarray}
is not necessarily satisfied. Therefore, eq.(\ref{PI3}) cannot be used, and thus 
there is no way to obtain the energy eigenvalues of the corresponding system from 
the path integral method. This indicates that eq.(\ref{PI1}) should 
normally contain some infinity in the integrations of the left hand side.

\section{Sommerfeld quantization and boundary condition}

As we see in the Introduction, the Sommerfeld quantization scheme is quite 
successful for obtaining the energy eigenvalue in some of the potential problems. 
Here, we briefly review the derivation of the Sommerfeld quantization rule 
from the Schr\"odinger equation in terms of the semiclassical approximation. 

\subsection{Sommerfeld quantization rule}

We start from the Schr\"odinger equation in one dimension
\begin{eqnarray}
\left[ -{ \hbar^2\over 2m}{\partial^2\over \partial x^2} 
+V(x) \right] \psi(x)=E\psi(x)  . 
\end{eqnarray}
Now we assume the wave function in the following shape
\begin{eqnarray}
\psi(x)=A e^{i{S\over \hbar}}
\end{eqnarray}
and expand the $S$ in terms of $\hbar$ as  $S=S_0+\hbar S_1 +\cdots $. By 
inserting the $S$ into the Schr\"odinger equation, we obtain the $S_0$ 
to the lowest order of $\hbar$
\begin{eqnarray}
{dS_0(x) \over dx}=\pm \sqrt{2m(E-V(x))}.
\end{eqnarray}
Therefore, the wave function $\psi(x)$ becomes
\begin{eqnarray}
\psi(x)=A e^{\pm{i\over \hbar} \int \sqrt{2m(E-V(x))}dx}. 
\end{eqnarray}
At this point, we require that the wave function should be a single-valued 
function, and therefore obtain the following constraint equation 
\begin{eqnarray}
\oint pdx \equiv \oint \sqrt{2m(E-V(x))}dx = 2\pi \hbar n, \ \ \ \ \ 
(n: \ {\rm integer \ or \ half \ integer}) 
\label{SQ1}
\end{eqnarray}
which is just the Sommerfeld quantization rule.

\subsection{Boundary condition in the Sommerfeld quantization}

The constraint equation of eq.(\ref{SQ1}) is obtained from the requirement 
that the wave function should be a single-valued function. 
This should not necessarily correspond to a boundary condition at infinity, 
but should be an important physical imposition on the state vector, and therefore 
we can obtain the energy spectrum from this condition. However, it should be noted 
that the  Sommerfeld quantization scheme can reproduce the right description of 
the  energy eigenvalue only if the system has some semiclassical nature like 
the Coulomb potential system. 

The boundary condition at large $x$ should reflect the quantum effect 
in the wave function of the corresponding particle.  This can be viewed 
from the uncertainty law that the large $x$ corresponds normally to the small 
momentum regions in which the quantum effect becomes most important. 
\vspace{0.2cm}
\\
\noindent
\textbullet \ {\bf Coulomb potential ($V(r)=-{Ze^2\over r}$) : } \ 
In the Coulomb potential, the strong part should be found in the  small $r$ region, 
and therefore the large momentum region is most important, and thus it can be 
considered to be almost semiclassical. Indeed, the Sommerfeld quantization scheme 
can give the right answer for the energy spectrum of the Coulomb potential system. 
We show the calculated result which is given in Introduction
\begin{eqnarray}
\oint pdr = 2\sqrt{2m|E|} \int_0^{{Ze^2\over |E|}} 
\sqrt{ {{Ze^2\over |E|}-r\over r} } dr  =nh \ \ \ \ \ 
\Longrightarrow \ \ \ \ \ E=-{m (Ze^2)^2\over 2 \hbar^2 n^2}.
\end{eqnarray}
The boundary condition of eq.(\ref{BC1}) is effectively included in the above 
equation. This can be seen since the integration is limited to the finite region 
of space coordinate which is physically acceptable area of space. 
\vspace{0.2cm}
\\
\noindent
\textbullet \ {\bf Harmonic oscillator potential ($V(x)={1\over 2}m\omega^2 x^2$) : } \ 
\begin{eqnarray}
\oint pdx \equiv 
\oint \sqrt{2m(E-V(x))}dx 
=2m\omega \oint_{-\sqrt{2E\over m\omega^2}}^{\sqrt{2E\over m\omega^2}}
\sqrt{{2E\over m\omega^2}-x^2}dx=2m\omega \, {2E\over m\omega^2}\, {n\over 2}
= 2\pi \hbar n . 
\end{eqnarray}
Thus, by considering the fact that the value of $n$ 
in eq.(\ref{SQ1}) can be an integer as well as a half integer, we obtain 
the energy $E$
\begin{eqnarray}
E=\hbar \omega \left(n+{1\over 2}\right), \ \ \ \ \ \ (n=0,1,2,\cdots). 
\end{eqnarray}
\vspace{0.2cm}
\\
\noindent
\textbullet \ {\bf Square well potential ($V(x)=-V_0 \ \mbox{for} \ |x|<a, 
\ \ \mbox{otherwise} \ \ V(x)=0$) : } \ 
\begin{eqnarray}
\oint pdx \equiv 
\oint \sqrt{2m(E-V(x))}dx =2\sqrt{2m(E+V_0)}2a = 2\pi \hbar n . 
\end{eqnarray}
Thus, we obtain 
\begin{eqnarray}
E=-V_0+{\hbar^2\, ({\pi n\over 2})^2\over 2m a^2}.  \label{sqw}
\end{eqnarray}
This can be compared with the result obtained from the quantum mechanics 
calculation for the square well potential model. This energy cannot be given 
in terms of the analytical expression. Instead, one should solve the following 
equation for $E$
\begin{eqnarray}
\alpha &=& \sqrt{{2m\left(E+V_0\right)a^2\over \hbar^2}}, \ \ \ 
\beta = \sqrt{-{2mEa^2\over \hbar^2}} \\ 
\beta &=& \alpha \tan \alpha. \label{sqw2}
\end{eqnarray}
Approximate solutions can be found when $\alpha \simeq {\pi\over 2}n$ which 
should be expected from $\tan \alpha \simeq 0$. In this case, 
we can immediately obtain the energy $E$ as
\begin{eqnarray}
E \simeq -V_0+{\hbar^2\, ({\pi n\over 2})^2\over 2m a^2}
\end{eqnarray}
which is just the same as the semiclassical result of eq.(\ref{sqw}). 
In this case, however,  eq.(\ref{sqw2}) is not necessarily 
satisfied, and thus the result of the bound state energy should  be far from 
reliable. 
\vspace{0.2cm}
\\
\noindent
\textbullet \ {\bf $\delta$ function potential ($V(x)=-V_\delta \delta(x)$) : } \ 
The limitation of the Sommerfeld quantization rule can be well exhibited if one 
solves the energy spectrum of the  $\delta$ function potential with the semiclassical 
method. The answer is that there is no way to express the energy $E$ in terms of 
$V_\delta$. This is partly because of the mathematical difficulty due to 
the generalized function of $\delta(x)$ in the square root, but mainly because of 
the limitation of the Sommerfeld quantization rule itself. 
This difficulty is better understood if we calculate the energy spectrum 
of the  $\delta$ function potential from the square well potential result. 
In this case, we start from eq.(\ref{sqw2}) 
\begin{eqnarray}
\sqrt{-{2mEa^2\over \hbar^2}}=\sqrt{2m(E+V_0)a^2\over \hbar^2}\, 
\tan \sqrt{2m(E+V_0)a^2\over \hbar^2}.  \label{del}
\end{eqnarray}
Here, we make $a \rightarrow 0, \ \ V_0 \rightarrow \infty$, but keep 
$V_\delta =2aV_0 $ finite. Therefore, eq.(\ref{del}) can be 
written to a good approximation as
\begin{eqnarray}
\sqrt{-2mE\over \hbar^2} \simeq \sqrt{mV_\delta\over \hbar^2 }\, 
\sqrt{mV_\delta\over \hbar^2 }   \label{del2}
\end{eqnarray}
which leads to the energy $E$
\begin{eqnarray}
E=-{mV_\delta^2\over 2 \hbar^2}
\end{eqnarray}
and this is indeed a correct energy eigenvalue of the $\delta$ function potential 
in quantum mechanics problem. 

As we note above, the energy of the square well potential from the Sommerfeld 
quantization rule is obtained without making use of eq.(\ref{sqw2}) 
while the energy of the $\delta$ function potential is obtained only if we make 
use of eq.(\ref{sqw2}). 
From this discussion, we see that the 
Sommerfeld quantization rule cannot be applied to the $\delta$ function potential.

\section{Summary}

We have clarified the similarity and difference between the path 
integral formulation and the Sommerfeld quantization rule. 
The similarity of the two methods is concerned with the classical 
mechanics since both approaches start from the equations of classical mechanics. 

On the other hand, there is a significant difference between them. 
The Sommerfeld quantization rule can properly take into account 
the boundary condition which is crucial for solving the bound state 
problem. On the other hand, the path integral method cannot include 
any of the wave function information, and therefore it cannot 
consider the boundary condition at infinity. Therefore, the path 
integral method can only solve the harmonic oscillator potential 
problem in which there is no need of the boundary condition. 

The investigation of the present paper may not contain any new physics 
except that the limitation of the path integral method is made clear 
for the first time. In this respect, it should present an important step 
for readers to realize that any numerical calculations based on the path 
integral formulation cannot give the right energy spectrum of physically 
interesting potential models. 

In addition, it is shown that the Sommerfeld quantization rule can be quite 
useful for calculating the spectrum of some quantum mechanics models, but 
at the same time, we explicitly present the limitation of the semiclassical 
picture in terms of the square well potential calculation.


\end{document}